\documentclass[10pt,letterpaper]{article}
\usepackage[top=0.85in,left=2.75in,footskip=0.75in]{geometry}

\usepackage{amsmath,amssymb}

\usepackage{changepage}

\usepackage[utf8x]{inputenc}

\usepackage{textcomp,marvosym}

\usepackage{cite}

\usepackage{nameref,hyperref}

\usepackage[right]{lineno}

\usepackage{microtype}
\DisableLigatures[f]{encoding = *, family = * }

\usepackage[table]{xcolor}

\usepackage{array}

\newcolumntype{+}{!{\vrule width 2pt}}

\newlength\savedwidth

\raggedright
\usepackage[aboveskip=1pt,labelfont=bf,labelsep=period,justification=raggedright,singlelinecheck=off]{caption}

\makeatletter
\renewcommand{\@biblabel}[1]{\quad#1.}
\makeatother

\usepackage{lastpage,fancyhdr,graphicx}
\usepackage{epstopdf}
\fancyheadoffset[L]{2.25in}
\fancyfootoffset[L]{2.25in}
\DeclareMathOperator{\diag}{diag}
\DeclareMathOperator{\tr}{tr}
\DeclareMathOperator{\supp}{supp}
\newcommand{\suchthat}{ \,\, \big | \,\, }
\renewcommand{\L}{\mathcal{L}}
\renewcommand{\a}{\mathfrak{a}}
\newcommand{\na}{\bar{\a}}
\usepackage{tikz}
	\usetikzlibrary{calc}
	\usetikzlibrary[topaths]
\usepackage{bm}
\usepackage{color}
	\definecolor{dkgreen}{rgb}{0,0.5,0}
	\definecolor{dkred}{rgb}{0.6,0.0,0}
	\definecolor{dkblue}{rgb}{0,0.0,0.6}
	\definecolor{gray}{rgb}{0.5,0.5,0.5}
	\definecolor{mauve}{rgb}{0.58,0,0.82}
	\definecolor{magenta}{rgb}{1,0,.5}

\newtheorem{expl}{Example}[section]
\newtheorem{rem}{Remark}[section]

\begin{document}
\vspace*{0.2in}

\begin{flushleft}
{\Large
\textbf\newline{Community structure detection and evaluation during the pre- and post-ictal hippocampal depth recordings} % Please use "sentence case" for title and headings (capitalize only the first word in a title (or heading), the first word in a subtitle (or subheading), and any proper nouns).
}
%short title: CA1 region of the hippocampus and community structure determination
\newline
\\
Keivan {Hassani Monfared}\textsuperscript{1*\Yinyang},
Kris Vasudevan\textsuperscript{1\Yinyang},
{Jordan S.} Farrell\textsuperscript{2\ddag},
{G. Campbell} Teskey\textsuperscript{2\ddag}
\\
\bigskip
\textbf{1} Department of Mathematics and Statistics, University of Calgary, Calgary, Alberta T2N 1N4, Canada
\\
\textbf{2} Hotchkiss Brain Institute, Cumming School of Medicine, University of Calgary, Calgary, AB T2N 1N4, Canada
\\
\bigskip

% Insert additional author notes using the symbols described below. Insert symbol callouts after author names as necessary.
% 
% Remove or comment out the author notes below if they aren't used.
%
% Primary Equal Contribution Note
\Yinyang These authors contributed equally to the final draft of the manuscript.

% Additional Equal Contribution Note
% Also use this double-dagger symbol for special authorship notes, such as senior authorship.
\ddag These authors designed and conducted the experiment, and also contributed equally to the final draft of the manuscript.

%\textcurrency Current Address: Dept/Program/Center, Institution Name, City, State, Country % change symbol to "\textcurrency a" if more than one current address note
% \textcurrency b Insert second current address 
% \textcurrency c Insert third current address


% Group/Consortium Author Note
%\textpilcrow Membership list can be found in the Acknowledgments section.

% Use the asterisk to denote corresponding authorship and provide email address in note below.
* k1monfared@gmail.com

\end{flushleft}

	\section*{Abstract}
		Detecting and evaluating regions of brain under various circumstances is one of the most interesting topics in computational neuroscience. However, the majority of the studies on detecting communities of a functional connectivity network of the brain is done on networks obtained from coherency attributes, and not from correlation. This lack of studies, in part, is due to the fact that many common methods for clustering graphs require the nodes of the network to be `positively' linked together, a property that is guaranteed by a coherency matrix, by definition. However, correlation matrices reveal more information regarding how each pair of nodes are linked together. In this study, for the first time we simultaneously examine four inherently different network clustering methods (spectral, heuristic, and optimization methods) applied to the functional connectivity networks of the CA1 region of the hippocampus of an anaesthetized rat during pre-ictal and post-ictal states. The networks are obtained from correlation matrices, and its results are compared with the ones obtained by applying the same methods to coherency matrices. The correlation matrices show a much finer community structure compared to the coherency matrices. Furthermore, we examine the potential smoothing effect of choosing various window sizes for computing the correlation/coherency matrices.

\section*{Author summary}
	The CA1 region of the hippocampus is known to be a critical component in the memory function of the brain. Moreover, elicited and spontaneous seizures in the CA1 region of hippocampus are known to cause EEG state changes that may be related to long-lasting changes in oxygen levels. Since the local field potential (LFP) depth recordings with a 16-contact point depth electrode at all stages of the ictal-period in the CA1 region are amenable to detailed analysis, we use a graph-theoretic approach to focus on an interesting problem of community structures 	in neuronal network of this region to study the pre-ictal and post-ictal states. Our	results yield convincing answers to the role of differing community structures during the pre-ictal and the post-ictal periods. We prefer the use of correlation matrices to coherency matrices to understand the subtle and strongly-varying community structures in the post-ictal period in the case of an anesthetized rat.

\section{Introduction}
	Understanding the neuronal behaviour in the brain under different contexts is made possible with the advent of a mathematical model that uses the theory of complex networks \cite{bullmoresporns09}. Such brain networks fall into either structural connectivity or functional connectivity systems \cite{bullmoresporns09, meunierlambiottebullmore10}. A common thread in these systems is the inherent modularity \cite{meunierlambiottebullmore10}.  Furthermore, there is a hierarchical organization of modularity in brain networks \cite{meunierlambiottebullmore10}.  A question on the presence of hierarchical organization of modularity arises when one considers a localized region of the brain such as the hippocampus. Here, a node of the network refers to a contact point of an electrode recording the local field potential (LFP) of a subpopulation of neurons. 

	One application of the abovementioned analysis is to understand how the brain changes during and after seizures. A long-lasting period of hypoxia occurs in the postictal period and could be associated with changes in community structure across electrode contacts \cite{farrelletal2017,farrelletal16}. To establish if there are noticeable characteristics to be found in electroencephalogram (EEG) recordings during the pre-ictal and post-ictal periods, measurements were made on anesthetized rats under specific experimental conditions. The data were collected at a sampling rate of $1000$Hz for a period of $117$ minutes. Here we display sample data from one rat before, during, and after a seizure.

\begin{figure}[h]
		\begin{center}
		\includegraphics[width=\linewidth]{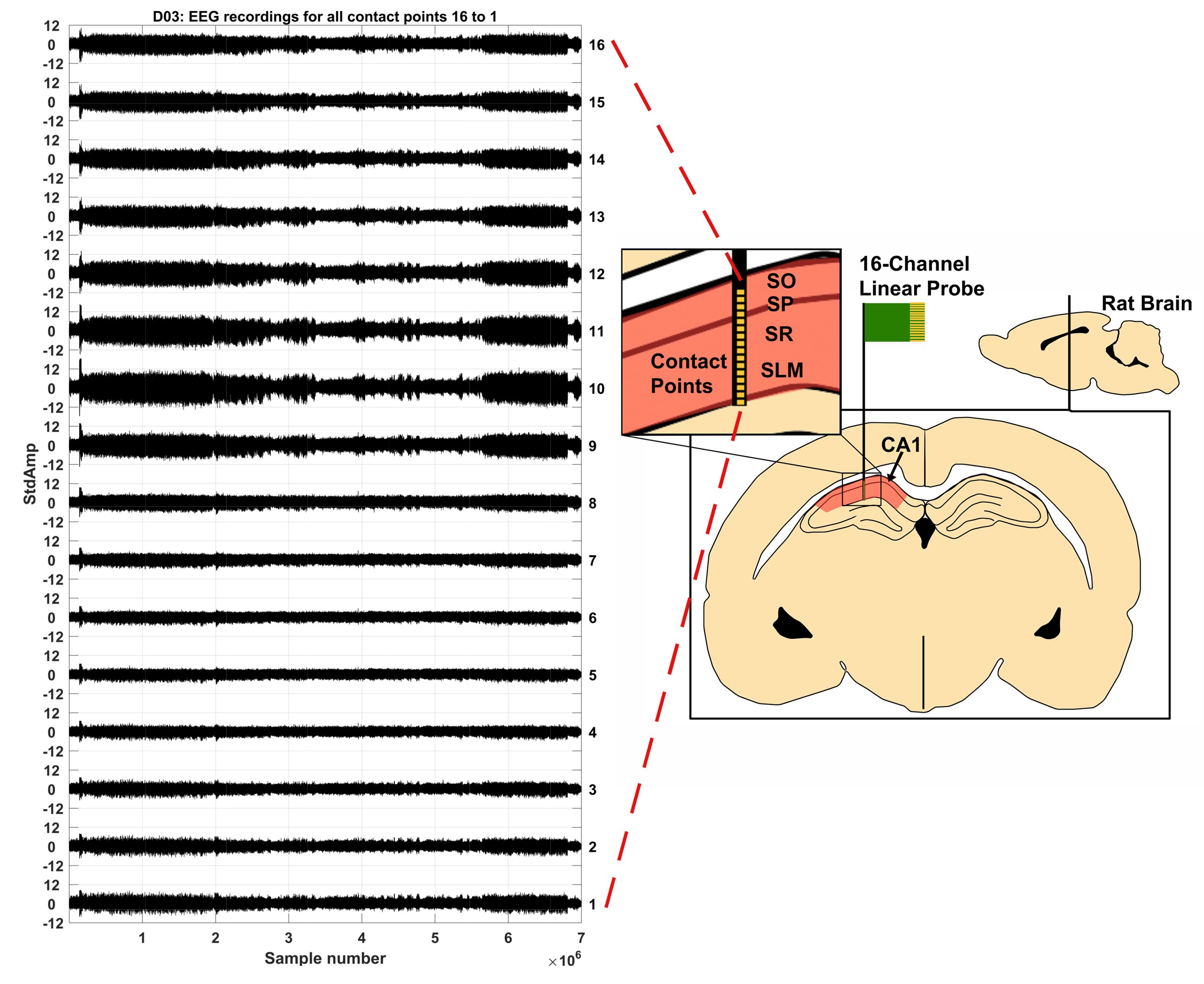}
		\end{center}
	\caption{All $16$ time-series from the contact points of the $16$-point depth electrode. The
electrode contacts are in different layers of CA1, namely stratum oriens (SO), stratum
pyramidale (SP), stratum radiatum (SR) and stratum lacunosum moleculare (SLM).} \label{fig1}
	\end{figure}

\begin{figure}[h]
		\begin{center}
			\includegraphics[scale=.3]{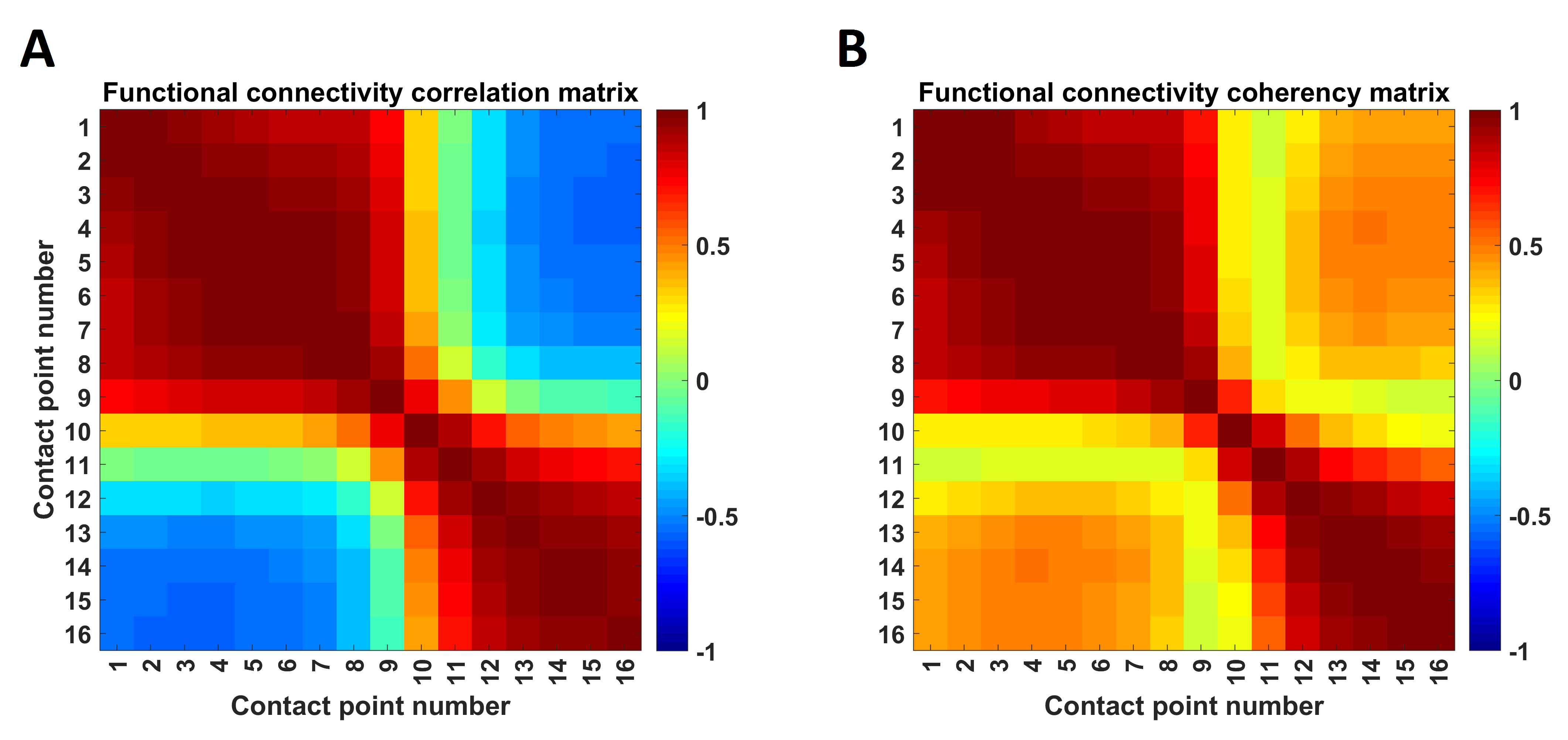}
		\end{center}
	\caption{A correlation (A) and a coherency (B), functional connectivity matrix corresponding to the windowed time series of D$03$ at the pre-ictal period. Row and column indices correspond to contact point number.} \label{fig2}
	\end{figure}

	It is common to study the networks arising from the coherency of the signals recorded at each node of the electrode to analyze the modularity structure of the brain \cite{Bassettetal11}. The novelty of this paper is to also analyze the networks obtained by considering cross-correlation matrices, and to compare the results obtained from coherency. Hence, we processed the resulting LFP time-series, created sliding-window subsets of data and carried out zero-lag cross-correlation and coherency on them.  We call the resulting windowed correlation matrices functional connectivity matrices of the CA1 hippocampal region. In Fig \ref{fig1} we show the time-series from all contact points of the $16$-point depth electrode. The contact points are in different layers of CA1 namely stratum oriens, stratum pyramidale, stratum radiatum, stratum lacunosum moleculare. As stated earlier, each contact point is a node in the cell layer and the windowed functional correlation connectivity matrix is construed as reflecting the interaction of both the local network and the different projections of synaptic inputs to CA1, arriving from other brain regions such as CA3 of hippocampus, entorhinal cortexm septal nuclei etc. In Fig \ref{fig2} we include examples of a windowed correlation functional connectivity matrix (left) and of a windowed coherency functional connectivity matrix (right).

	One interesting feature that emerges from this example is that the correlation matrix yields a signed graph (corresponding to the left matrix in Fig \ref{fig2}). We also find that a modular structure is associated with this graph.  Signed graphs have drawn interest in recent years among graph theorists \cite{atayhua16, bronskideville14, bullmoresporns09, chung97, gomezetal09, guozhao12, guozhang16, harary53, lili09, zhanetal17} in the context of both static and dynamic behaviour of the graphs.

	Brain networks on a global scale are known to undergo changes with time both in healthy brains and also, brains with certain natural or induced disorders \cite{burnsetal14}.  We ask whether or not there are changes with time in the functional connectivity matrices obtained for anesthetized rats under different conditions.  Furthermore, we are interested in finding out the role of the modular structure of signed graphs with time.  Also, we need to know if there is a hierarchical organization of the modular structure at a smaller scale on the brain, that is, the CA1 region of the hippocampus, in comparison with well-established hierarchical modularity embedded in brains \cite{meunierlambiottebullmore10}.

\begin{figure}[h]
		\begin{center}		
		\includegraphics[width=\linewidth]{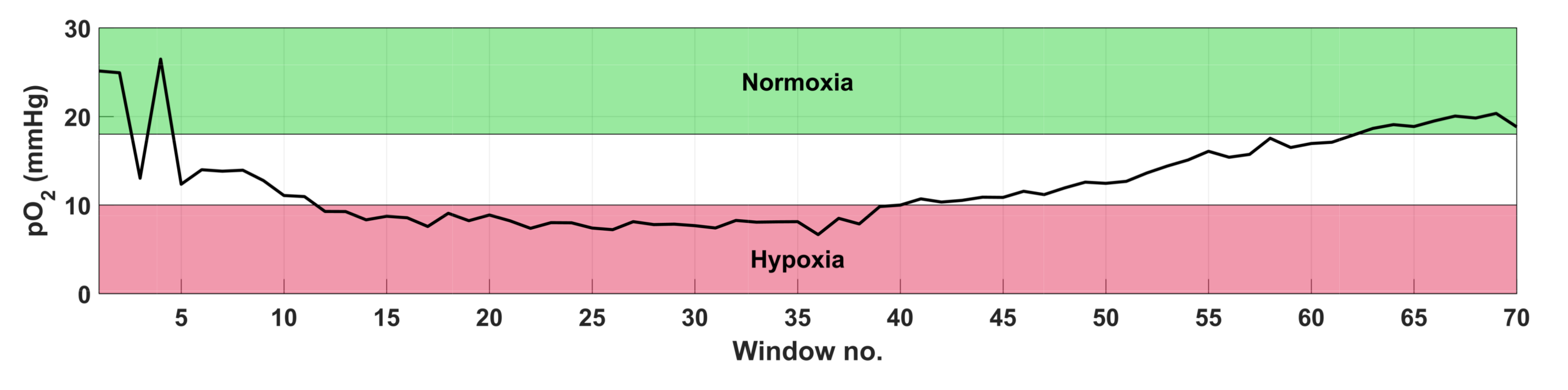}
		\end{center}
	\caption{A simulated oxygen level profile for the CA$1$ region of the hippocampus of the rat after seizure.} \label{fig3}
	\end{figure}

	To answer these questions in this paper, we present here four mathematical methods, namely the Fiedler method \cite{fiedler75}, the Girvan-Newman method \cite{newmangirvan04}, the spectral coordinates method \cite{wuwululi17}, and an adaptation of simulated annealing method to maximize the signed modularity \cite{gadelkarimetal12, gomezetal09}, to obtain reasonable clusterings of signed graphs. All four methods are unsupervised, in the sense that the number of clusters is not given as an input, and it is obtained by maximizing a cost function in the algorithm. We compare the results obtained from the four methods to stress the importance of modularity in the evolving functional connectivity structure. Also, we stress the impact of the cross-correlated window size on the community structure and modularity indices. We consider as well the coherency as a measure instead of the cross-correlation coefficient for the same size windows to illustrate the role of unsigned coherency graphs. Finally, we speculate on our present results in their relevance to the oxygen profiles in post-ictal hypoxia studies (see Fig \ref{fig3}).

\section{Methods}
\subsection{Signed graphs and the corresponding matrices}
	Functional connectivity matrices arising out of LFP data recorded in the CA1 region of the hippocampus with an electrode containing 16 depth-points are examples of undirected signed graphs. We are interested in community detection in such graphs.  Community detection in networks has been extensively explored for graphs which are unsigned \cite{fortunato10, gadelkarimetal14, gadelkarimetal12, girvannewman02, newman01, newman06}, that is, the edges (the nonzero off-diagonal elements of functional connectivity matrices) are positive.  The recorded data we have chosen as an example can be divided into non-overlapping windowed data with each window made up from a fixed number of data samples. 
	Here, we considered four sets of windows for the data and used $5000$, $10000$, $20000$, and $100000$ samples for each window in each set.  We show in Fig \ref{fig2} a selected windowed correlation and a selected windowed coherency functional connectivity matrix. The figure reveal not only positive edges or positive correlation coefficients but also negative edges or negative correlation coefficients in some windows of the correlation matrices. We define here the ratio of the number of negative edges to the total number of edges as anticorrelation index.  Community detection in networks or graphs with a non-zero anticorrelation index has drawn considerable interest in the last few years \cite{lili09, wuwululi17, gomezetal09, zhanetal17}.

	To detect communities in an unsigned graph, several spectral graph methods are currently in use (see the survey \cite{fortunato10}). One such method is the Fiedler method \cite{fiedler75}. This method gets less and less accurate as the graph gets more and more connected, as is expected. Also, it cannot deal with signed graphs in its simple form, due to the use of the Laplacian matrix which is defined for unsigned graphs. Girvan and Newman \cite{girvannewman02} introduced an optimization method which uses an objective or cost function which we henceforth call as modularity. % Intrinsically, this allows one to make a comparison of different communities as a result of graph partitioning. 
	Despite the success of the modularity approach of Newman and Girvan  \cite{newmangirvan04}, its applicability to signed graphs being relevant to the present data was explored with a modified form of the modularity expression \cite{gomezetal09}.  In view of the importance of negative correlation in both functional properties of the networks and non-linear dynamics on such networks \cite{bronskideville14}, we undertake not only a study of both spectral graph methods such as Fiedler eigenvector method and spectral coordinates approach but also an examination of optimization methods using modified form of the modularity index initially defined by Girvan and Newman \cite{girvannewman02}.  In the next section, we describe the methods used.  Also, we present the approach to look at the hierarchy of the modular network. In general, for small graphs we suggest using the simulated annealing method that maximizes the modularity. It is also possible to use the result of either Fiedler, Girvan-Newman, or spectral coordinates method as the starting point in the search space of the simulated annealing method.

	Throughout the paper, we consider only real symmetric matrices, $I_n$ denotes the identity matrix of size $n$, and if the size is clear from the context, we write $I$. Our model is based on graphs that represent correlation matrices. A correlation matrix has values between $-1$ and $1$, is symmetric, and all diagonal entries are equal to $1$ (every node correlates with itself). 	
	
	\begin{expl} \label{exacorrelationmatrix}
		Here is a real symmetric matrix with positive and negative entries:
		\begin{equation}
			A = \frac{1}{10} \left(10 I + \left[ \begin{array}{rrrrr|rrrr}
				\color{gray}0 & \color{dkred}1 & \color{gray}0 & \color{dkred}8 & \color{dkblue}-2 & \color{gray}0 & \color{gray}0 & \color{gray}0 & \color{dkblue}-1\\ 
				\color{dkred}1 & \color{gray}0 & \color{dkred}6 & \color{dkred}7 & \color{dkred}3 & \color{gray}0 & \color{gray}0 & \color{dkred}8 & \color{gray}0\\ 
				\color{gray}0 & \color{dkred}6 & \color{gray}0 & \color{dkred}5 & \color{dkblue}-6 & \color{gray}0 & \color{gray}0 & \color{gray}0 & \color{gray}0\\ 
				\color{dkred}8 & \color{dkred}7 & \color{dkred}5 & \color{gray}0 & \color{dkred}1 & \color{gray}0 & \color{dkblue}-7 & \color{gray}0 & \color{gray}0\\ 
				\color{dkblue}-2 & \color{dkred}3 & \color{dkblue}-6 & \color{dkred}1 & \color{gray}0 & \color{gray}0 & \color{gray}0 & \color{gray}0 & \color{gray}0\\ \hline
				\color{gray}0 & \color{gray}0 & \color{gray}0 & \color{gray}0 & \color{gray}0 & \color{gray}0 & \color{dkred}7 & \color{dkred}9 & \color{dkblue}-1\\ 
				\color{gray}0 & \color{gray}0 & \color{gray}0 & \color{dkblue}-7 & \color{gray}0 & \color{dkred}7 & \color{gray}0 & \color{dkred}6 & \color{dkred}5\\ 
				\color{gray}0 & \color{dkred}8 & \color{gray}0 & \color{gray}0 & \color{gray}0 & \color{dkred}9 & \color{dkred}6 & \color{gray}0 & \color{dkred}6\\ 
				\color{dkblue}-1 & \color{gray}0 & \color{gray}0 & \color{gray}0 & \color{gray}0 & \color{dkblue}-1 & \color{dkred}5 & \color{dkred}6 & \color{gray}0 
			\end{array} \right]\right).		
		\end{equation}
	\end{expl}
	
	First, we introduce some graph theoretical terminology, then we assign a graph to each correlation matrix. The tuple $G = (V,E,W,\Sigma)$ is called a signed weighted undirected graph (with no parallel edges, but potentially with loops) with vertices $V = \{1,2,\ldots,n\}$, edges $E = \big\{ e_1 = \{i_1,j_1\} ,e_2 = \{i_2,j_2\},\ldots,e_m = \{i_m,j_m\} \big\}$, edge weights $W = \{ w_1, w_2, \ldots, w_m \}$, and signs of edges $\Sigma = \{ \sigma_1, \sigma_2, \ldots, \sigma_m \}$. Here we assume that $0 \leq w_i \leq 1$ and $\sigma_i = \pm 1$, for all $i = 1,2,\ldots,m$.
	
	The (weighted) adjacency matrix of $G$ is defined to be an $n\times n$ real symmetric matrix $A = A(G) = \begin{bmatrix}
		a_{ij}
	\end{bmatrix}$, where $a_{ij} = w_{\ell}$ (with $e_\ell = \{i,j\}$), whenever vertex $i$ is adjacent to vertex $j$, and zero otherwise. Note that for an unweighted graph, all $w_i = 1$, hence $A$ becomes the classical adjacency matrix of $G$. For each vertex $i$ let $d_i$ denote the degree of vertex $i$, that is, 
	\begin{equation}
		d_i = \sum_{i\sim j} w_j = \sum_{j=1}^{n} a_{ij}.
	\end{equation}
	Similarly, the signed (weighted) adjacency matrix of $G$ is an $n\times n$ real symmetric matrix $S = S(G) = \begin{bmatrix}
		s_{ij}
	\end{bmatrix}$, where $s_{ij} = \sigma_{\ell} w_{\ell}$, with $e_\ell = \{i,j\}$. 
	
	\begin{expl} \label{exsignedgraphofacorrelationmatrix}
	The graph representing matrix $A-\frac{1}{10} I$ of Example \ref{exacorrelationmatrix} is shown in Fig $\ref{fig4}$.	
	
	\begin{figure}[h]
	\begin{center}
		\includegraphics[scale=.2]{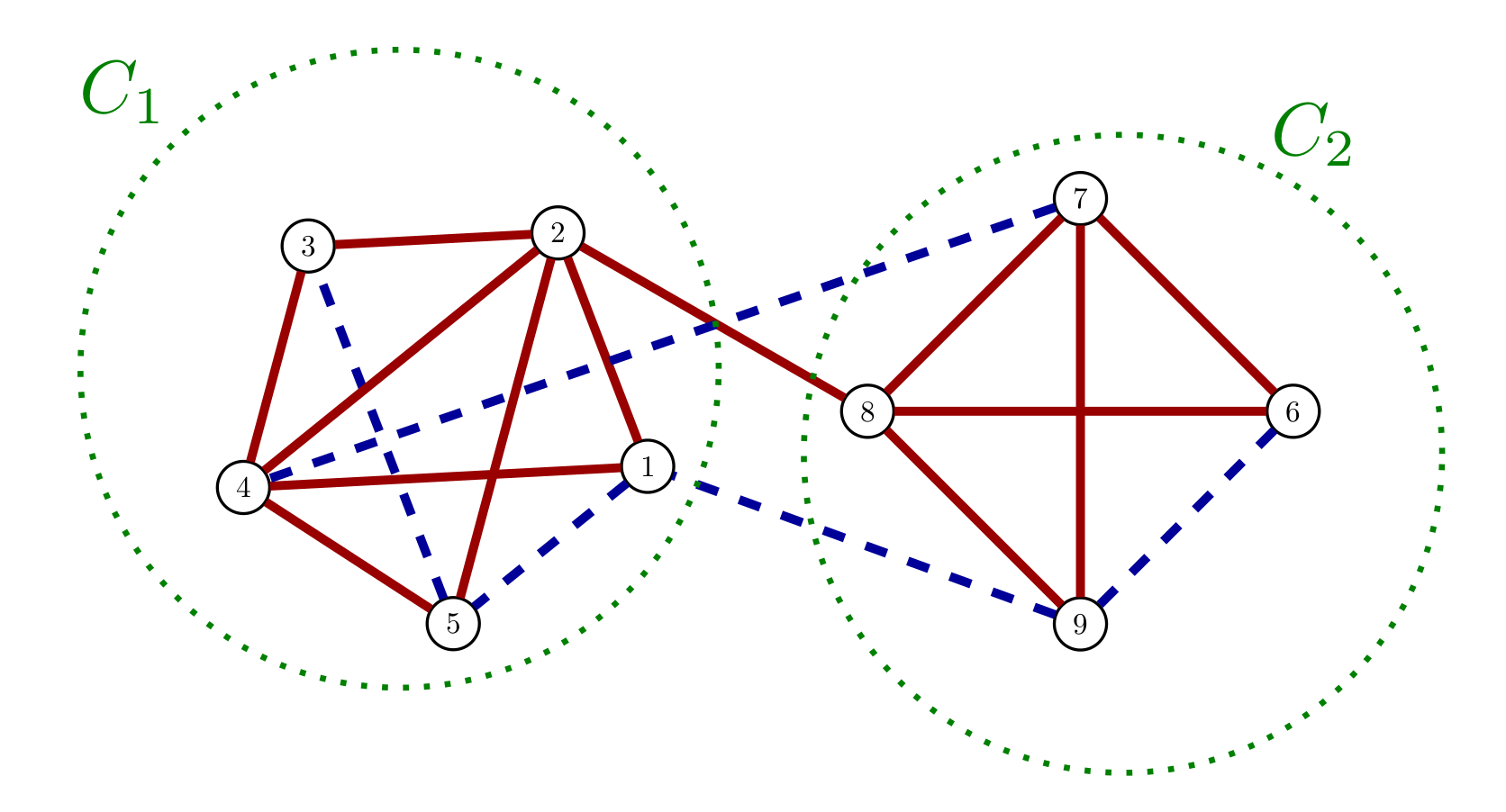}
	\end{center}
	\caption{A signed graph with two apparent clusters. Straight ({\color{dkred}red}) lines represent positive edges, and dashed  ({\color{dkblue}blue}) lines are negative edges. The dotted ({\color{dkgreen}green}) bubbles represent two clusters.}\label{fig4}
	\end{figure}
	\end{expl}
	
	Define $D = \diag(d_1, d_2, \ldots, d_n)$ to be the diagonal matrix with $d_i$'s down the main diagonal. Then the Laplacian of $G$ is defined to be $L = D - A$, and if $G$ does not have an isolated vertex its normalized Laplacian is defined to be $\L = D^{-\frac{1}{2}} L D^{-\frac{1}{2}}$.
	
	\begin{expl}
		The adjacency matrix and the Laplacian matrix of the \emph{unsigned} (weighted) graph of Example \ref{exsignedgraphofacorrelationmatrix} respectively are
		\begin{equation}			
			A = \frac{1}{10} \left[ \begin{array}{rrrrr|rrrr}
			 \color{gray}0 &  1 &  \color{gray}0 &  8 & 2 &  \color{gray}0 &  \color{gray}0 & \color{gray}0 & 1\\		
			 1 &  \color{gray}0 &  6 &  7 & 3 &  \color{gray}0 &  \color{gray}0 & 8 &  \color{gray}0\\	
			 \color{gray}0 &  6 &  \color{gray}0 &  5 & 6 &  \color{gray}0 &  \color{gray}0 & \color{gray}0 &  \color{gray}0\\
			 8 &  7 &  5 &  \color{gray}0 & 1 &  \color{gray}0 &  7 & \color{gray}0 &  \color{gray}0\\
			 2 &  3 &  6 &  1 & \color{gray}0 &  \color{gray}0 &  \color{gray}0 & \color{gray}0 &  \color{gray}0\\ \hline
			 \color{gray}0 &  \color{gray}0 &  \color{gray}0 &  \color{gray}0 & \color{gray}0 &  \color{gray}0 &  7 & 9 &  1\\
			 \color{gray}0 &  \color{gray}0 &  \color{gray}0 &  7 & \color{gray}0 &  7 &  \color{gray}0 & 6 &  5\\
			 \color{gray}0 &  8 &  \color{gray}0 &  \color{gray}0 & \color{gray}0 &  9 &  6 & \color{gray}0 &  6\\
			 1 &  \color{gray}0 &  \color{gray}0 &  \color{gray}0 & \color{gray}0 &  1 &  5 & 6 & \color{gray}0\\
			\end{array} \right],
		\end{equation}		
		\begin{equation}			
			L = \frac{1}{10} \left[ \begin{array}{rrrrr|rrrr}
			 \bm 12 &  -1 &  \color{gray}0 &  -8 & -2 &  \color{gray}0 &  \color{gray}0 & \color{gray}0 & -1\\		
			 -1 &  \bm {25} &  -6 &  -7 & -3 &  \color{gray}0 &  \color{gray}0 & -8 &  \color{gray}0\\	
			 \color{gray}0 &  -6 &  \bm {17} &  -5 & -6 &  \color{gray}0 &  \color{gray}0 & \color{gray}0 &  \color{gray}0\\
			 -8 &  -7 &  -5 &  \bm {28} & -1 &  \color{gray}0 &  -7 & \color{gray}0 &  \color{gray}0\\
			 -2 &  -3 &  -6 &  -1 & \bm {12} &  \color{gray}0 &  \color{gray}0 & \color{gray}0 &  \color{gray}0\\ \hline
			 \color{gray}0 &  \color{gray}0 &  \color{gray}0 &  \color{gray}0 & \color{gray}0 &  \bm {17} &  -7 & -9 &  -1\\
			 \color{gray}0 &  \color{gray}0 &  \color{gray}0 &  -7 & \color{gray}0 &  -7 &  \bm {25} & -6 &  -5\\
			 \color{gray}0 &  -8 &  \color{gray}0 &  \color{gray}0 & \color{gray}0 &  -9 &  -6 & \bm {29} &  -6\\
			 -1 &  \color{gray}0 &  \color{gray}0 &  \color{gray}0 & \color{gray}0 &  -1 &  -5 & -6 & \bm {13}\\
			\end{array} \right].
		\end{equation}
	\end{expl}

\subsection{Community structure and modularity index}		
	A (non-overlapping) clustering of a graph $G=(V,E)$ is a partitioning $C$ of the vertex set $V$ of $G$. That is, to construct $C = \{ V_1, V_2, \ldots, V_k \}$ such that \begin{itemize}
		\item $V = \displaystyle\bigcup_{i=1}^{k} V_i$,
		\item $V_i \cap V_j = \emptyset$, for all $i\neq j$, and 
		\item $V_i \neq \emptyset$, for all $i = 1,2,\ldots, k$.
	\end{itemize}
		
	For a given clustering $C = \{ V_1, V_2, \ldots, V_k \}$, an edge $e = \{u,v\} \in E$ with $u \in V_i$ and $v \in V_j$ is called \begin{itemize}
		\item an intra edge, if $i = j$, and
		\item an inter edge, if $i \neq j$.
	\end{itemize}
	Intuitively, (in a simple graph) a clustering of a graph $G$ is reasonable in one sense, if the number of inter edges is relatively small compared to the number of intra edges. That is roughly speaking, the induced subgraph on each $V_i$ is `dense', and the subgraph of $G$ obtained by removing all the intra edges is `sparse'. In a signed graph, one often wants that the induced subgraph on each $V_i$ to be dense with positive edges and sparse with negative edges, and the subgraph of $G$ obtained by removing all the intra edges is sparse with positive edges and/or dense with negative edges. In this section we review a few common methods to cluster a graph, and measures to quantify how meaningful the clustering is. In general settings one of the most common approaches to clustering is the agglomerative (bottom-up) approach, which starts with each vertex as one cluster, and then pairs of most similar clusters are merged together, iteratively. This approach is not suggested when the community structure of the network is known \cite{newmangirvan04}. On the other hand, one can use a divisive (top-down) approach which starts with one cluster containing all the vertices, and then recursively, each cluster is split into two clusters. All of the four methods considered in this paper are divisive (top-down) methods. See page $5$ of \cite{gadelkarimetal14} and references therein for discussions on bottom-up and top-down methods.
		
	\subsection{Method A. Using the Fiedler vector of an unsigned graph for clustering it via its Laplacian matrix}
		Consider the (normalized) Laplacian matrix of a connected graph $G$. Note that if $G$ is not signed, then both $L$ and $\L$ are diagonally dominant matrices, and hence they are positive semidefinite. That is, all of the eigenvalues are nonnegative. It can be shown that the nullity (i.e. the number of zero eigenvalues) of both of these matrices is equal to the number of connected components of graph $G$  \cite{chung97}. Also, it can be shown that the second smalles eigenvalue of $\L$ lie in the interval $[0,\frac{n}{n-1}]$ \cite{chung97}. Moreover, since the row sums of $L$ are all equal to zero, for each connected component there is a ``signature'' eigenvector corresponding to it. That is, if $G$ has $k$ connected components $C_1, C_2, \ldots, C_k$, then for each $i = 1,2,\ldots, k$ the vector $\bm x_i$ below is an eigenvector corresponding to the eigenvalue zero of $L$. 
		\begin{equation}
			\bm x_i = \begin{bmatrix}
				x_{i,1} & x_{i,2} & \cdots & x_{i,n}
			\end{bmatrix}^\top,
		\end{equation}
		with $x_{i,j} = 1$ if and only if vertex $j$ is in component $C_i$, and zero otherwise. The second smallest eigenvalue of $L$ is called the algebraic connectivity $\a$ of $G$ \cite{fiedler73}. Intuitively, the smaller $\a$ is, the more of a meaningful clustering does $G$ have into two clusters. Fiedler has shown that the two clusters can be obtained by considering an eigenvector $\bm x = \begin{bmatrix}
			x_1 & x_2 & \cdots & x_n
		\end{bmatrix}^\top$ corresponding to this eigenvalue \cite{fiedler75}. Namely, the two clusters are $V_1 = \{ v \in V \suchthat x_i \leq 0 \}$ and $V_2 = \{ v \in V \suchthat x_i > 0 \}$. This can be used in an iterative process to chop each cluster into two pieces, in order to get more clusters. Keeping track of the steps, this yields a hierarchical clustering of the graph. One concern here is that even if the graph is highly connected (e.g. a complete graph) Fiedler's method provides two clusters, since the second eigenvector always will have both positive and negative entries for a connected graph. Hence, we need a criteria to stop/postpone the iterative process for a given highly connected subgraph, usually well before chopping a graph on $n$ vertices into $n$ clusters of single vertices.
		
		A natural choice for the stopping/postponing criteria is the algebraic connectivity of the graph. If it is ``very large'', then we stop the process. However, the second smallest eigenvalue of the Laplacian of a graph (i.e. the algebraic connectivity) depends on the number of vertices, and ``very large'' becomes subjective. In order to avoid this dependency on the number of vertices, one might consider the second smallest eigenvalue of the normalized Laplacian of the graph, which we call it the normalized algebraic connectivity $\na$ of $G$. The parameter $\na$ is ``less'' dependent on the number of vertices of $G$ and has the same properties as $\a$ that we are interested in. Namely, $\na$ is zero if and only if the graph is disconnected%, and it is nondecreaasing in number of edges. 
		. What makes $\na$ a better choice than $\a$ is the following property:
		\begin{itemize}
			\item If $G$ is not the complete graph, then $\na \leq 1$.
		\end{itemize}
		This will provide us with a reasonable criteria on when to stop chopping the graph into two pieces, namely, when $\na$ is relatively ``large''. In practice, we iterate the process of dividing each cluster into two clusters until all the vertices are separated from each other. Hence the way that $\na$ is used is as a postponing criteria. That is, at each iteration, the normalized algebraic connectivities of all current clusters are compared, and the iteration is done on a cluster with minimum $\na$. This reveals a meaningful hierarchy at the end.

	\subsection{Method B. Spectral coordinates of a signed graph for clustering it via its signed adjacency matrix}
		Let $S$ be the signed adjacency matrix of a signed weighted graph, and let $\bm x_1, \bm x_2, \ldots, \bm x_n$ be eigenvectors of $S$ corresponding to eigenvalues $\lambda_1 \geq \lambda_2 \geq \cdots \geq \lambda_n$. Normalize $\bm x_i$'s so that they all have unit 2-norm, and put them in a canoncial form where their first nonzero entry is positive. Now, define $\bm u_i = (u_{i,1}, u_{i,2}, \ldots, u_{i,n})$ where $u_{i,j} = x_{j,i}$. We call each $u_i$ the spectral coordinate of vertex $i$ \cite{wuwululi17}.
		
		\begin{equation}
		\begin{array}{rc}
		& 	\begin{array}{cccccc}
				\bm x_1\phantom{.} & \bm x_2\phantom{.} & \cdots & \bm x_k\phantom{.} & \cdots & \phantom{.}\bm x_n \\
				\downarrow & \downarrow & & \downarrow & & \downarrow
			\end{array} \\
		\begin{array}{r}
			\\
			\\
			\\
			\bm u_i \rightarrow \\
			\\
			\\
		\end{array} & 
			\begin{array}{cccc|cc}
				x_{11} & x_{21} & \cdots & x_{k1} & \cdots & x_{n1} \\
				x_{12} & x_{22} & \cdots & x_{k2} & \cdots & x_{n2} \\
				\vdots & \vdots & \ddots & \vdots & \ddots & \vdots \\
				x_{1i} & x_{2i} & \cdots & x_{ki} & \cdots & x_{ni} \\
				\vdots & \vdots & \ddots & \vdots & \ddots & \vdots \\
				x_{1n} & x_{2n} & \cdots & x_{kn} & \cdots & x_{nn} \\
			\end{array}
		\end{array}
		\end{equation}
		
		For a fixed $k$ we project $\bm u_i$ onto its first $k$ components 
		\begin{equation}
			\overline{\bm u_i} = (x_{1i}, x_{2i}, \ldots, x_{ki}).
		\end{equation}
		For a vector $\bm x$ its support, $\supp(\bm x)$ is the set of indices $i$ where $\bm x_i$ is nonzero
		\begin{equation}
			\supp(\bm x) = \{ i \suchthat \bm x_i \neq 0\}.
		\end{equation}
		The idea is that for a graph with $k$ connected components $V_1$, $V_2$, $\ldots$ ,$V_k$ with relatively equal sizes, if $A$ does not have repeated eigenvalues, then each of $k$ eigenvectors corresponding to the largest $k$ eigenvalues of $A$ have nonzero entries only on the vertices corresponding to one connected component, and their supports do not intersect. As a result, the spectral coordinates $\bm u_i$ of all vertices $i$ in a connected component $V_j$ lie on the $j$-th axis in a $k$ dimensional space. Now we can look at general graphs as perturbations of such a graph, and we expect that if we only add a few edges (specially in a `balanced' way), then the spectral coordinates of the vertices of each connected component still lie close to each other, and close to some axis. Then various methods (such as k-means) are employed to cluster these points in $\mathbb{R}^k$, and hence to cluster the vertices of the original graph.
		
		One technical issue that needs to be taken into account is that the spectral coordinates obtained from normalized eigenvectors with the canonical choice above still might not perfectly represent clusters of a graph even when the graph is disconnected. To illustrate this, note that when a graph is disconnected, the eigenvectors corresponding to $k$ largest eigenvectors might have positive or negative entries when they are nonzero. In this case they are still recognizable when we cluster the spectral coordinates according to what axis they lie on, regardless of which side of the origin (positive or negative) they lie on. But after perturbation this makes it hard (if not impossible) for clustering algorithms such as k-means to identify such points. In order to get around this technical problem, we scale $\bm u_i$'s so that their largest entry in magnitude is positive. 
		
		\begin{expl}
			Consider the disconnected graph in Fig $\ref{fig5}$ (left) and a perturbation of it (right). The spectral coordinates of the perturbed graph is shown in Fig $\ref{fig6}$.

			\begin{figure}[h]
			\begin{center}
				\includegraphics[width=\linewidth]{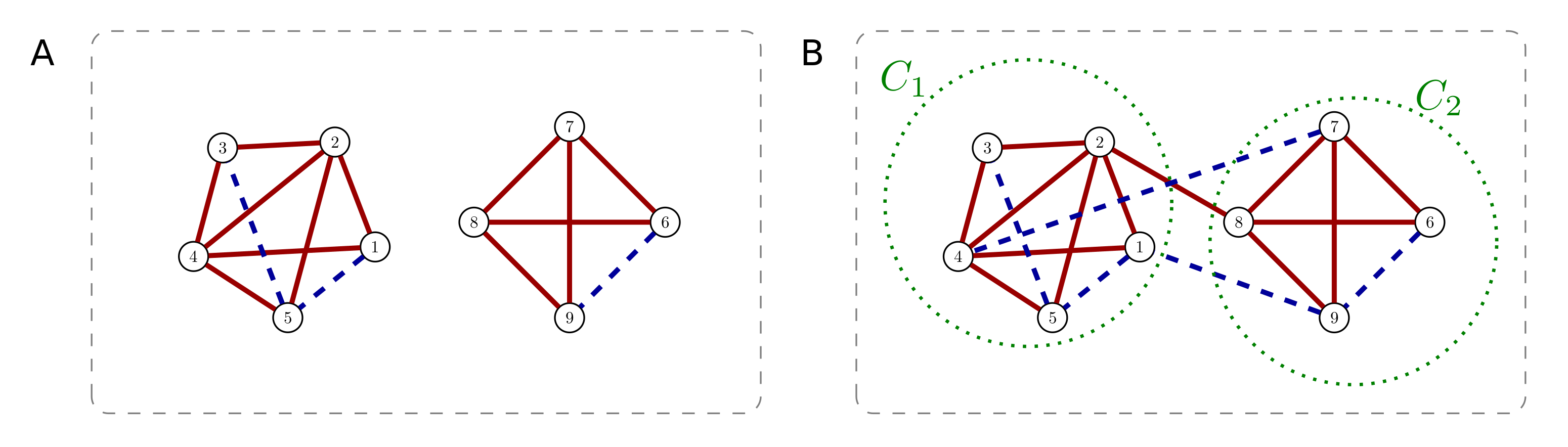}
			\end{center}
			\caption{Left: A disconnected graph with two connected components. Straight ({\color{dkred}red}) lines represent positive edges, and dashed  ({\color{dkblue}blue}) lines are negative edges. Right: A perturbation of the left graph which is now connected and with two apparent clusters}\label{fig5}
			\end{figure}

			\begin{figure}[h]
			\begin{center}
				\includegraphics[scale=.2]{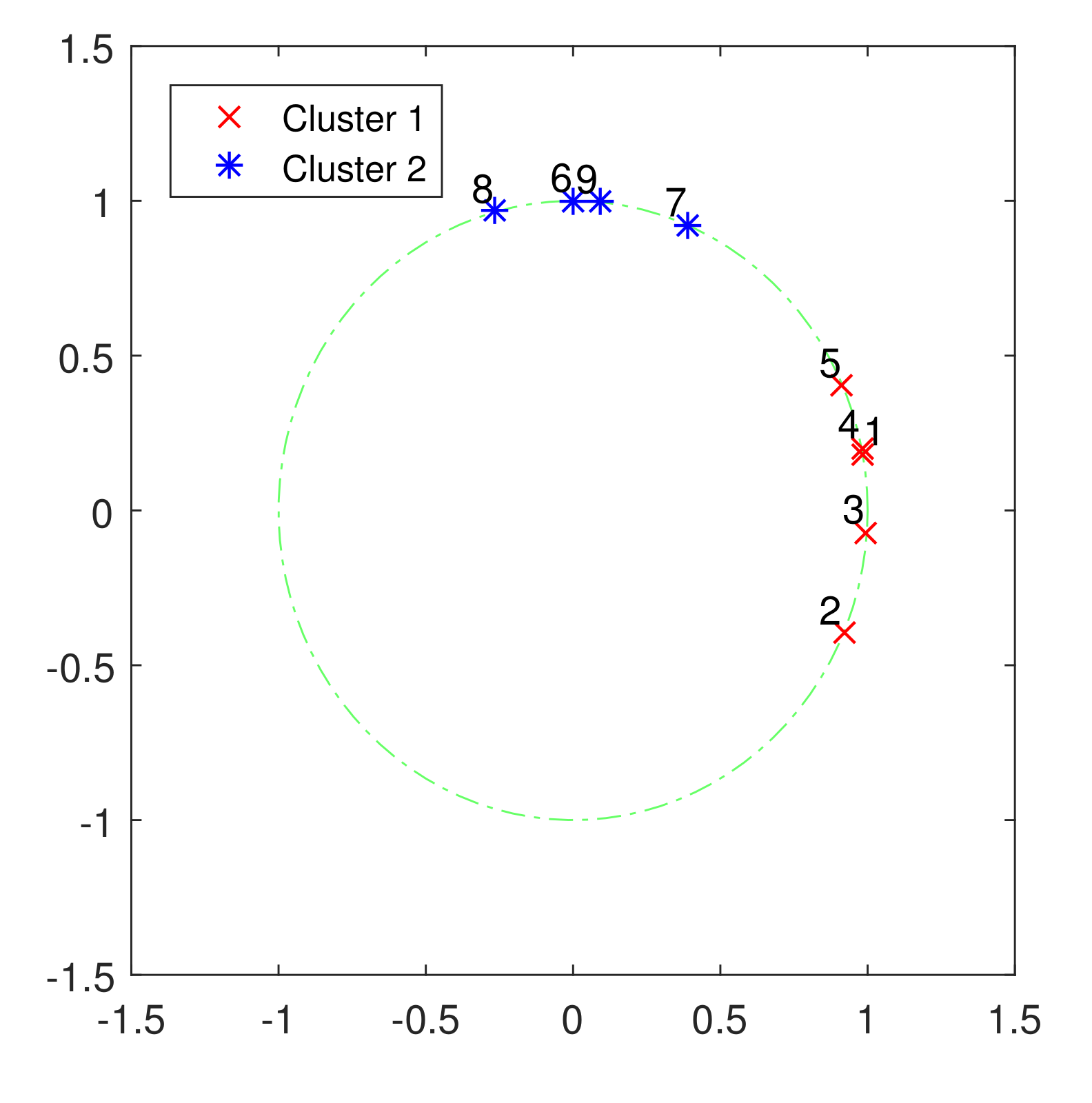}
			\end{center}
			\caption{The spectral coordinates (in $2$ dimensions) of the vertices of the right graph in Fig \ref{fig5} clustered into two clusters using k-means method. The green curve shows the unit circle for reference.}			
			\label{fig6}
			\end{figure}				
		\end{expl}

	\subsection{Method C. Girvan-Newman edge betweenness shortest path method for unsigned graphs}		
		Girvan and Newman \cite{girvannewman02} suggested the following concept of clustering. Each cluster is associated with a community structure. The clustering can be established by using the following strategy: 
		\begin{enumerate}
			\item Calculate the betweenness for all edges in the graph.
			\item\label{gvstep2} Remove the edge with the highest betweenness.
			\item\label{gvstep3} Recalculate betweenness for all edges affected by the removal.
			\item Repeat from step \ref{gvstep2} until no edges remain.
		\end{enumerate}
		The edge betweenness of an edge $e$ is the number of shortest paths between pairs of vertices that contain $e$. If there is more than one shortest path between a pair of vertices, each path is given equal weight such that the total weight of all of  the  paths  is  one \cite{newman01}. In step \ref{gvstep3} only edges that are in the same connected component as $e$ will be affected. 
		
	\subsubsection{Modularity index for general graphs}
		Aside from measures that try to describe how `well connected' a graph is, such as the (normalized) algebraic connectivity and the `clustering coefficient', there are various other measures quantifying how `good' a particular clustering of a graph is. That is, if a particular partitioning $C = \{ V_1, V_2, \ldots, V_k\}$ of the vertices of the graph is meaningful. One such measure given by Girvan and Newman in \cite{girvannewman02} is called the modularity of a clustering of the graph. Let $A$ be the adjacency matrix of a (weighted) unsigned graph $G$. Define a matrix  $E = \begin{bmatrix} e_{ij} \end{bmatrix}_{i,j=1}^k$, where $e_{ij}$ is the number of edges between vertices in $V_i$ and $V_j$ divided by the total number of edges of $G$, and let $a_i = \sum_{j=1}^{k} e_{ij}$. Note that when $i=j$ the edges in $V_i$ are counted only once. Then the (Girvan-Newman) modularity is defined as 
		\begin{equation}
			q = \sum_{i=1}^{k} e_{ii} - a_i^2 = \tr(E) - ||E^2||,
		\end{equation}
		where $||X||$ denotes the sum of all elements of $X$.
		%\footnote{
%		To clarify, let $A = \left[ \begin{array}{cc}
%			B & C \\ 
%			C^\top & D
%		\end{array} \right]$ be the adjacency matrix of a graph $G$, and let $b=||B||$, $c=||C||$, and $d=||D||$. Direct calculations show that for a clustering into two clusters that conforms to the block matrix above we have 
%		\begin{equation}
%			q = -2 \frac{4c^2 + (b+d)c - bd}{(b+2c+d)^2}.
%		\end{equation}
%		This implies that (since $b,c,d \geq 0$ and $c \leq \sqrt{bd}$)
%		\begin{equation}
%			q \leq 0 \text{ if and only if } c \geq \frac{2bd}{1+2(b+d)},
%		\end{equation}
%		where equality in the left hand side expression holds exactly when the equality in the right hand side expression holds. Note that when $bd = 0$, that is, when we have only one cluster, then $q = 0$. Furthermore, this shows that if the number of inter-edges grows as the number of intra-edges is fixed, then eventually the modularity becomes negative. 
%		}. 
		According to Newman and Girvan \cite{newmangirvan04}:
		
		 \begin{quotation}	
		 	 ``This quantity measures the fraction of the edges in the network that connect vertices of the same type (i.e., within-community edges) minus the expected value of the same quantity in a network with the same community divisions but random connections between the vertices. If the number of within-community edges is no better than random, we will get $q = 0$. Values approaching $q = 1$, which is the maximum, indicate strong community structure. In practice, values for such networks typically fall in the range from about $0.3$ to $0.7$. Higher values are rare.''
		\end{quotation}

		Note that the matrix computation with a weighted adjacency matrix will generalize the concept of modularity to weighted graphs \cite{newmangirvan04}. As we expect, the best clustering for a highly `clusterable' graph is given when all the edges are intra edges and there are no inter edges. This happens when the graph is disconnected where the connected components have the same number of edges, and the clusters are the connected components. When there are $k$ connected components, this yields modularity $1 - \frac{1}{k}$. On the other hand, the worst clustering of a graph is when all the edges are inter edges. This happens when the graph is a multi-partite graph, and the clusters are the parts. When the graph is bipartite this yields modularity $-2$.
		
		Now, consider a signed graph $G = (V,E,W,\Sigma)$. We define $G^+$ to be the graph obtained from $G$ by removing all the negative edges, and $G^-$ to be the graph obtained from $G$ by removing all the positive edges. Similarly, define $A^\pm$ to be the adjacency matrix of $G^\pm$. Note that both $A^+$ and $A^-$ are entrywise nonnegative. Furthermore, $A = A^+ + A^-$, and $S = A^+ - A^-$. Assume that $G^\pm$ has Girvan-Newman modularity $q^\pm$ and, and it has $m^\pm$ edges. Then the signed modularity of $G$ is defined to be \cite{gomezetal09}
		\begin{equation}\label{defsignedmodularity}
			q_s = \frac{m^+ q^+ - m^- q^-}{m^+ + m^-}.
		\end{equation}
		
		Note that $q_s$ coincides with $q$ when there are no negative edges. Furthermore, if a clustering $C$ of a graph $G = (V,E,W,\Sigma)$ yields a signed modularity $q$, then the same clustering of the graph $-G = (V,E,W,-\Sigma)$, the same graph with opposite signs yields a signed modularity $-q$. That is,
		\begin{equation}
			q_s(-G,C) = -q_s(G,C).
		\end{equation}
				
		\begin{rem}
			Note that $q_s$ is a convex combination of $q^+$ and $-q^-$. Considering the fact that $-2 \leq q^+,q^- < 1$, we conclude that $-2 \leq q_s \leq 2$, and the bounds are sharp.
		\end{rem}

		\subsection{Method D. Simulated annealing for maximizing signed modularity of signed graphs}		
			One might be interested in going over all possible clusterings of a signed graph and evaluate which one yields the maximum signed modularity. While this will find \emph{the} ``best'' clusterings with respect to the signed modularity, the search space grows exponentially fast. For example, for a graph on $16$ vertices, the number of all possible partitions is $10,480,142,147$ (that is, $B_{16}$, the $16$th Bell number).
			%from sage.combinat.expnums import expnums2; 
			%n = 16
			%expnums2(n+1, 1)[-1]
			
			One of the common ways to go around this huge search space, specifically when it is discrete, is to use probabilistic optimization methods such a simulated annealing \cite{gadelkarimetal14, gadelkarimetal12, kirkpatricketal83}. One of the benefits of the simulated annealing method is that the clustering works in a fixed number of steps which is defined by user. In our problem, the goal is to break down a given signed graph into two clusters such that the signed modularity of the clustering is maximum. Then we take each cluster and iterate the process on them separately in a hierarchical manner. In the best case (that is each iteration yields two equal size clusters) the size of the search space will be reduced to 
			\begin{equation}
				\sum_{k=0}^{\log_2(n)} \left( 2^{\frac{n}{2^k}-1} - 1 \right).
			\end{equation}
			%n = 16
			%L = [2^(n/(2^k) -1) -1 for k in range(log(n,2))]
			%sum(L)
			For example, when $n = 16$, this number is only $32,902$. And in the worst case (that is each iteration just separates one node from the rest) the size of the search space will be 
			\begin{equation}
				\sum_{k=0}^{n-1} \left( 2^{n-1-k} - 1 \right),
			\end{equation}
%			n = 16
%			L = [2^(n-k-1) -1 for k in range(n-1)]
%			sum(L)
			which for $n=16$ is $65,519$. In general, the \textbf{search space} is the set of all partitionings of the vertex set of size $n$ into two sets. The process starts from an \textbf{initial state}, that is a clustering into two sets, which we pick it to be random. The \textbf{objective function} is the signed modularity which is to be maximized. We need to define a `neighbour' for each state, which is a set of states (clusterings of the graph with two clusters) that can be reached from the current state in one `step'. Let us assume the current state is $C_0 = \{ V_1, V_2\}$, where $C_0$ is a clustering of the vertex set $V$ of the signed graph $G$ with a signed modularity $q_0$. That is, $V_1$ and $V_2$ are nonempty, $V_1 \cap V_2 = \emptyset$, and $V_1 \cup V_2 = V$. Furthermore, assume that $|V_1| > 1$. A \textbf{neighbour} of $C_0$ is $C_1 = \{V_3, V_4\}$, where $V_3 = V_1 \setminus \{v\}$ and $V_4 = V_2 \cup \{ v \}$, for a $v \in V_1$. In order to reach the neighbours, a random element of one of the clusters of size greater than one is chosen and put in the other cluster. This yields a search graph of diameter at most $n$. In our analysis we chose the number of steps to be $400$ for $n=16$. When $C_1$, a neighbour of $C_0$, is chosen, the signed modularity of it, $q_1$, is computed and compared to $q_0$. If $q_1 > q_0$, then we accept $C_1$ as the current state. Otherwise, $q_0 > q_1$, and $C_1$ is a worse clustering than $C_0$, but we accept $C_1$ as the current state with probability 
			\begin{equation}
				P(q_0,q_1,T) = e^{^{-\dfrac{q_0-q_1}{T}}},
			\end{equation}
			where $T$ is the `temperature'. The \textbf{temperature} is high at the beginning and reduces as the simulated annealing process progresses, resulting in less and less rates of acceptance of worse clusterings. At each temperature $500$ samples are taken to find the maximum. We stop the process when there are $8$ clusters. What we have is a list of $8$ clusterings of sizes $1$, $2$, $\ldots$, $8$,respectively. Then we find the maximum signed modularity among these $8$ clusterings. The clustering with maximum modularity is chosen to be the optimum clustering.

		\subsection{Coherency matrices and the use of above-mentioned methods}
			Similar to constructing $16\times 16$ correlation-windowed matrices, we used coherency as a measure to build $16\times 16$ coherency-matrices. Coherency, $C_{ij}^k$, is given by
			\begin{equation}
				C_{ij}^k = 
				\frac{{|P_{ij}^k|}^2}
				{P_{ii}^k P_{jj}^k{}'},
			\end{equation}
			where $P_{ij}^k$ is the cross-spectrum density between the $i$th and the $j$th contact point in the $16$-point electrode for a window $k$ of the sizes mentioned earlier.  Unlike the cross-correlation coefficients which lie between $-1$ and $1$, the coherency values are between $0$ and $1$. We show in Fig \ref{fig2}B an example of a coherency matrix corresponding to the same window we used for the correlation matrices. The window sampling followed the same set of values, $5000$, $10000$, $20000$, and $100000$ samples.

\section{Data Description}
\begin{figure}[h]
		\begin{center}
			\includegraphics[width=\linewidth]{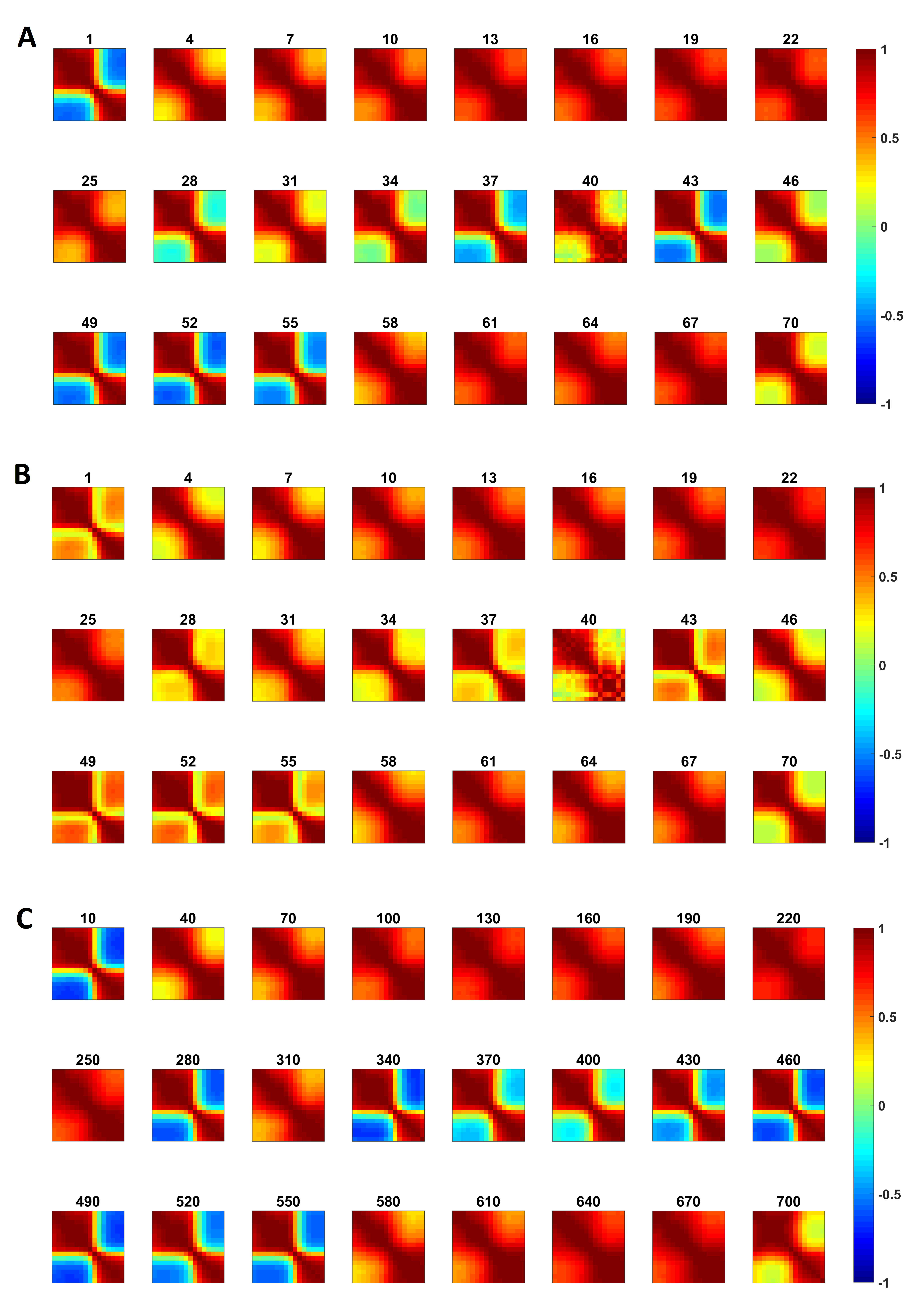} 
		\end{center}
		\caption{$24$ sample functional connectivity matrices of D$03$. A: Correlation, window size: $100000$. B: Coherency, window size: $100000$. C: Correlation, window size: $10000$.}
		\label{fig7}
	\end{figure}

	The data collected here are the $1000$Hz LFP's recorded at an electrode with $16$ contact points in the CA1 region of the
	hippocampus of an anesthetized rat that we refer to as D$03$. In this particular
	case, we recorded an approximately $4$ minute baseline before inducing a seizure by electrical kindling stimulation to the contralateral hippocampus \cite{goddard67}. The recording lasted for $117$ minutes, and we converted them to correlation and coherency
	matrices of order $16$ with $5000$, $10000$, $20000$, and $100000$ samples for each window, sufficiently large
	to understand the behaviour of LFP over a wide frequency
	spectrum. Fig \ref{fig7}A shows $24$ windows of 
	the functional connectivity matrices for D$03$. Each square represents a 
	$16\times 16$ matrix with entries between $-1$ and $1$, where the top left 
	corner entry (entry $(1,1)$) corresponds to vertex $1$, the bottom right entry 
	(entry $(16,16)$) corresponds to vertex $16$, and for example the top right 
	corner entry (entry $(1,16)$) corresponds to the weight and sign of the edge 
	$\{1,16\}$. Fig \ref{fig7}B corresponds to $24$ windows of the coherency matrices for D$03$.

\begin{figure}[h]
		\begin{center}
			\includegraphics[width=\linewidth]{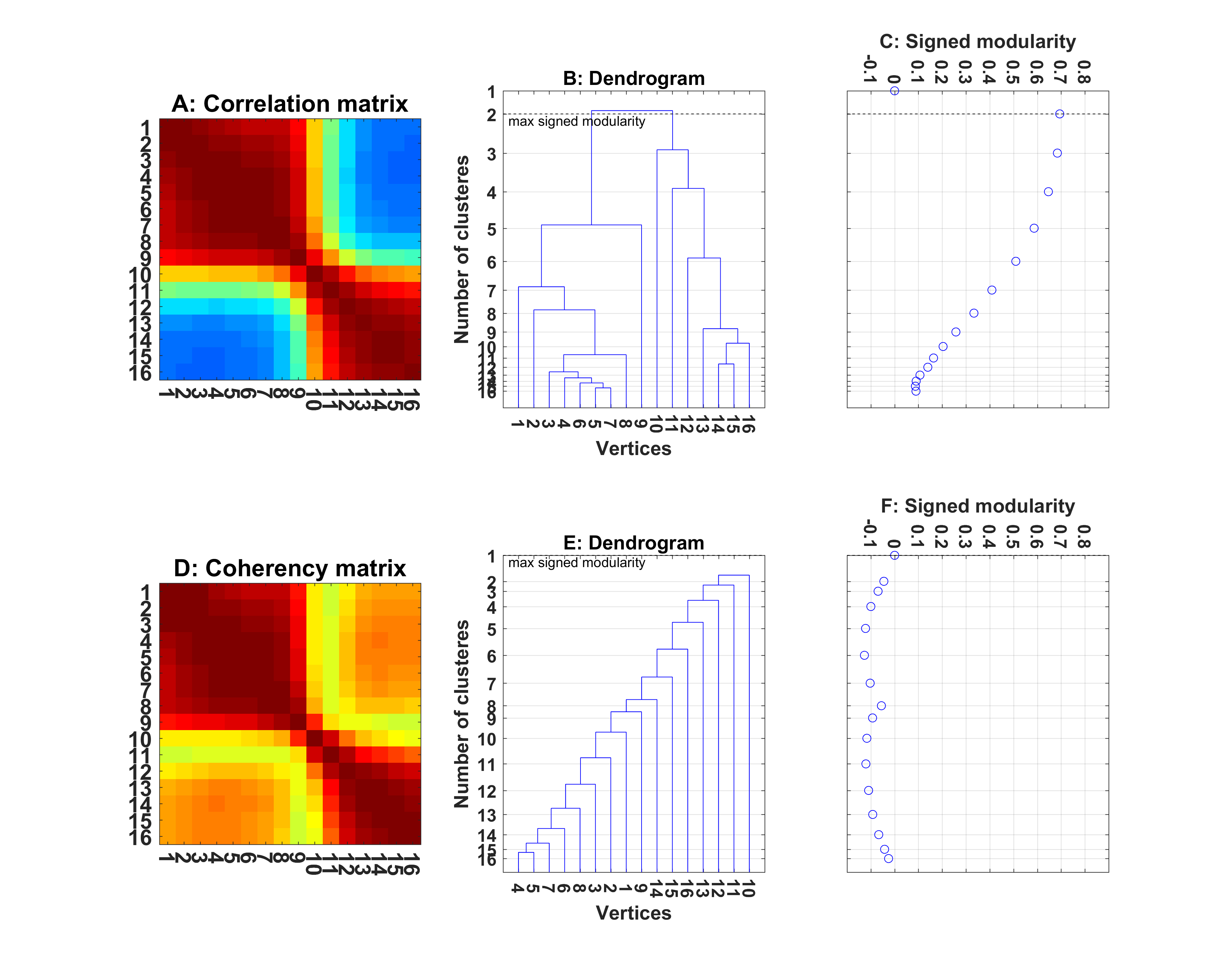}
		\end{center}
		\caption{A and D: The correlation and the coherency functional connectity matrices of order $16$ corresponding to a weighted (signed) graph representing the first window of D$03$. B and E: dendrograms representing the hierarchical clustering of the (signed) graphs of matrix in A and D, using Method D. C and F: the modularity of each clustering shown in B and E corresponding to various horizontal cuts of the dendrogram.}
		\label{fig8}
	\end{figure}

	In Fig \ref{fig8} a sample window (window $1$) of the correlation (top) and the coherency (bottom) functional connectivity matrices of D$03$ are shown, along with the dendrogram of a hierarchical clustering of them computed using Method D. Also, the signed modularity of the clusterings at each level is shown in the right, where the maximum modularity for them is marked.

\section{Results}
\begin{figure}[h]
		\begin{center}
			\includegraphics[width=\linewidth]{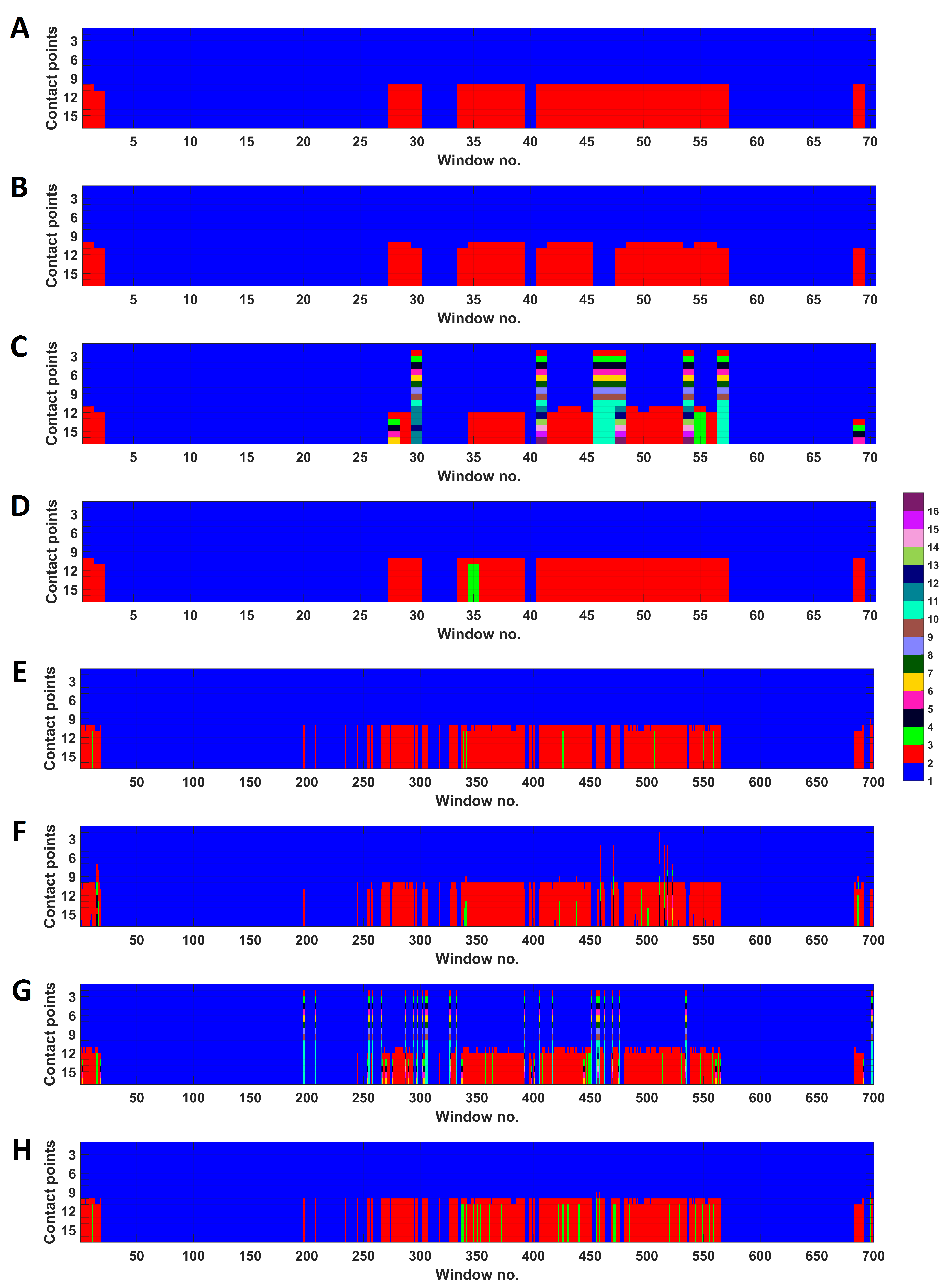}
		\end{center}
		\caption{The clusters for each window of the correlation matrices obtained by four methods. Each row in each panel is a contact point (a vertex of the grpah) and each column corresponds to one of the windowed functional connectivity matrices. Different clusters are distinguished with different colours. A and E: Method A, B and F: Method B, C and G: Method C, and D and H: Method D for window sizes $100000$, and $10000$ respectively.}\label{fig9}
	\end{figure}

	Our computations show that the signed graphs obtained from the correlation matrices for each windowed time series generally fall into two categories with some exceptions, in terms of the community structure: 
	\begin{enumerate}
		\item graphs which all of their vertices are highly connected to each other, and
		\item graphs with two highly segregated clusters where the vertices inside each cluster are highly connected.
	\end{enumerate}
	Almost all windows with two clusters show the vertices $1,2,\ldots,9$ are clustered together and vertices $10,11,\ldots,16$ are in the other cluster. The exceptions are when some algorithms for some window included vertex $10$ and/or $11$ in the other cluster, or identified it as its own cluster, as shown in Fig \ref{fig9}, as well as the results from method C which for some windows finds the optimal clustering with a large number of clusters. Further analysis of the latter exceptions shows that the modularity index of these signed networks increases only ever so slightly when there is a large number of clusters, compared to having only two or three clusters. Hence the unsupervised method prefers the large number of clusters, as it classifies it as a ``better'' clustering. However the signed modularity of the results of method C on those windows are lower than those of the other three methods as shown below.
	
	\begin{figure}[h]
		\begin{center}
			\includegraphics[width=\linewidth]{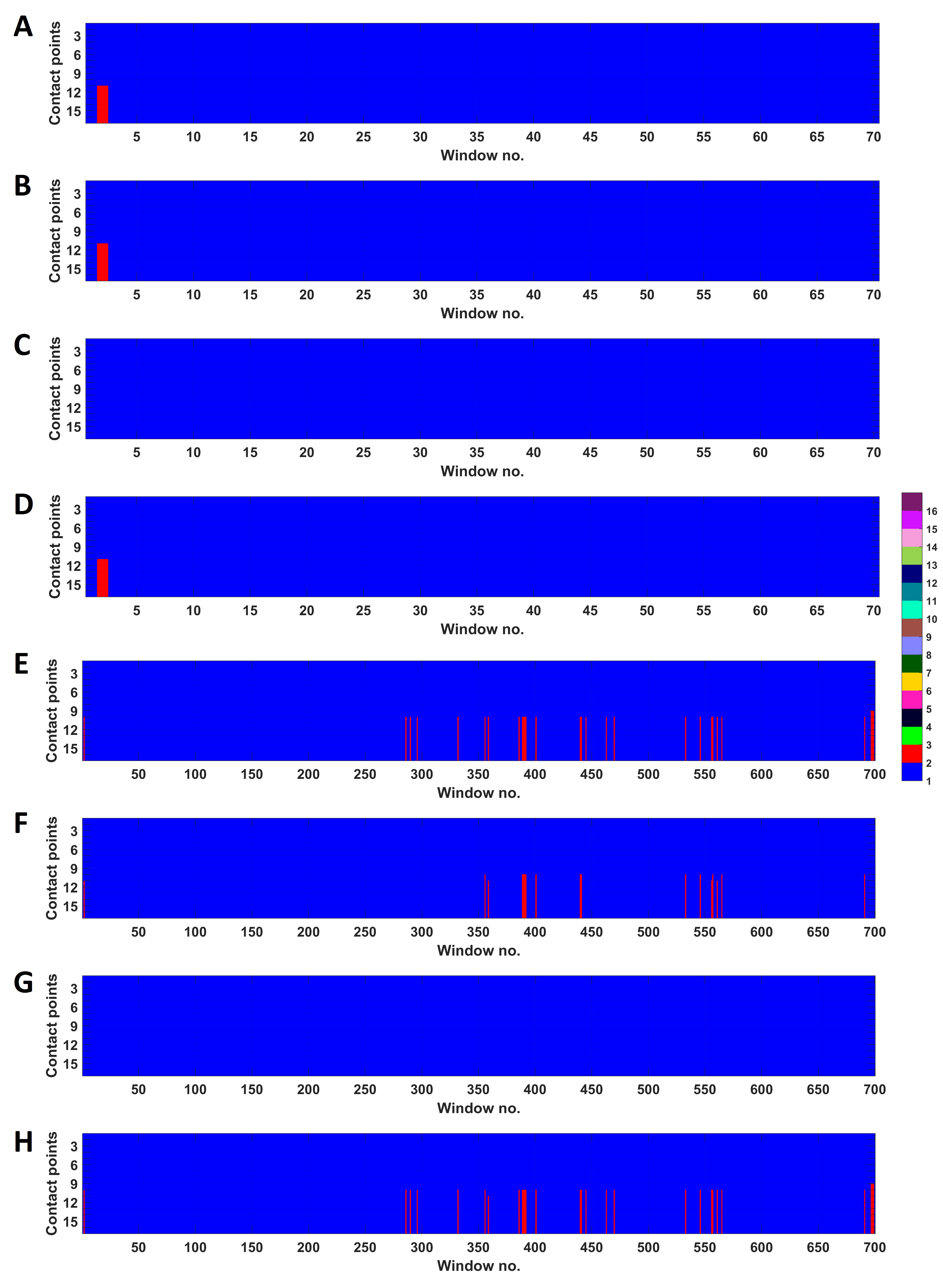}
		\end{center}
		\caption{The clusters for each window of the coherency matrices obtained by four methods. Each row in each panel is a contact point (a vertex of the graph) and each column corresponds to one of the windowed functional connectivity matrices. Different clusters are distinguished with different colours. A and E: Method A, B and F: Method B, C and G: Method C, and D and H: Method D for window sizes $100000$, and $10000$ respectively.}
		\label{fig10}
	\end{figure}
	
\begin{figure}[h]
		\begin{center}
			\includegraphics[width=\linewidth]{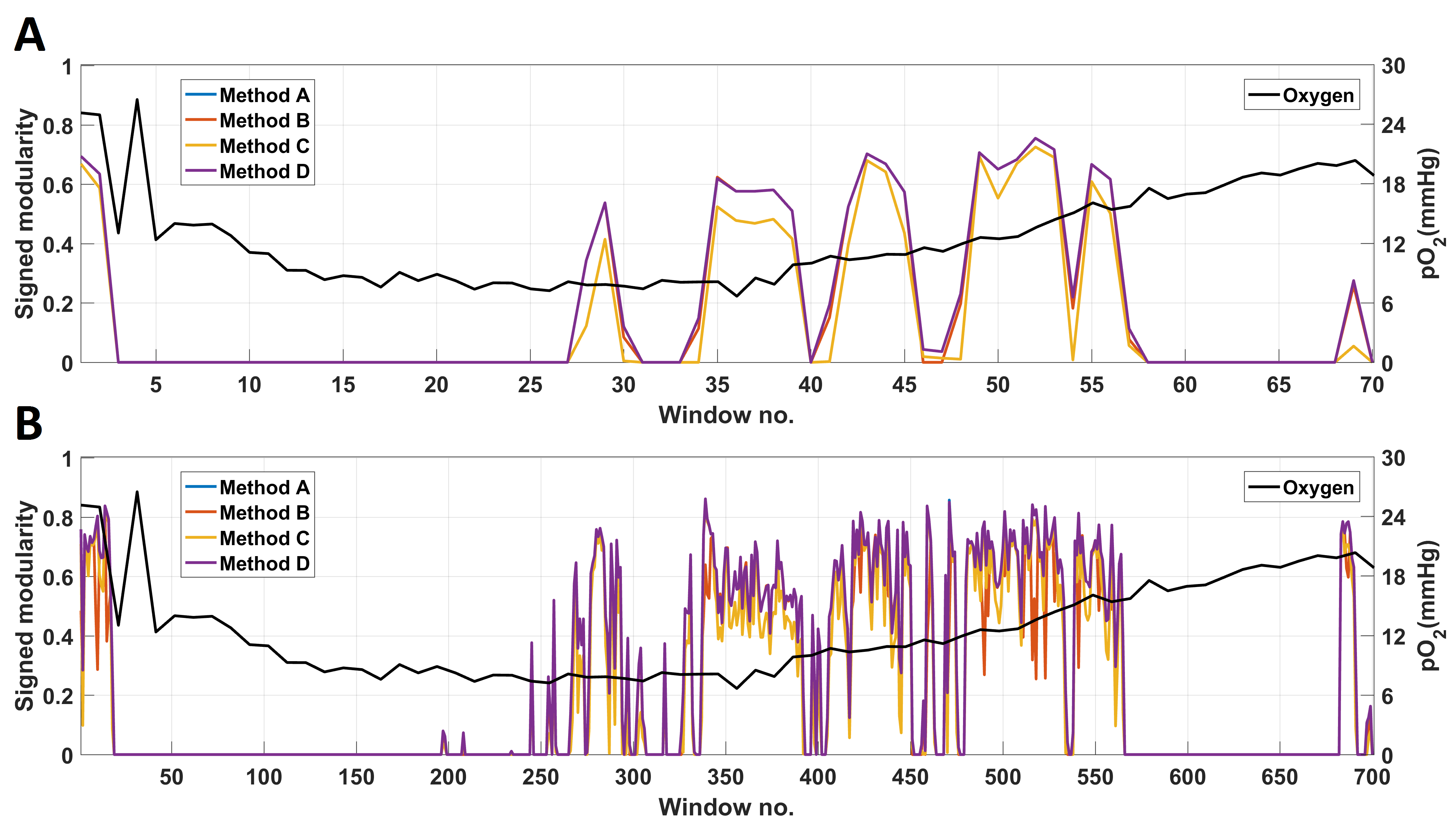}
		\end{center}
	\caption{The signed modularity obtained from all four methods for correlation matrices. A: window size: $100000$ and B: window size: $10000$.}
	\label{fig11}	
	\end{figure}

\begin{figure}[h]
		\begin{center}
			\includegraphics[width=\linewidth]{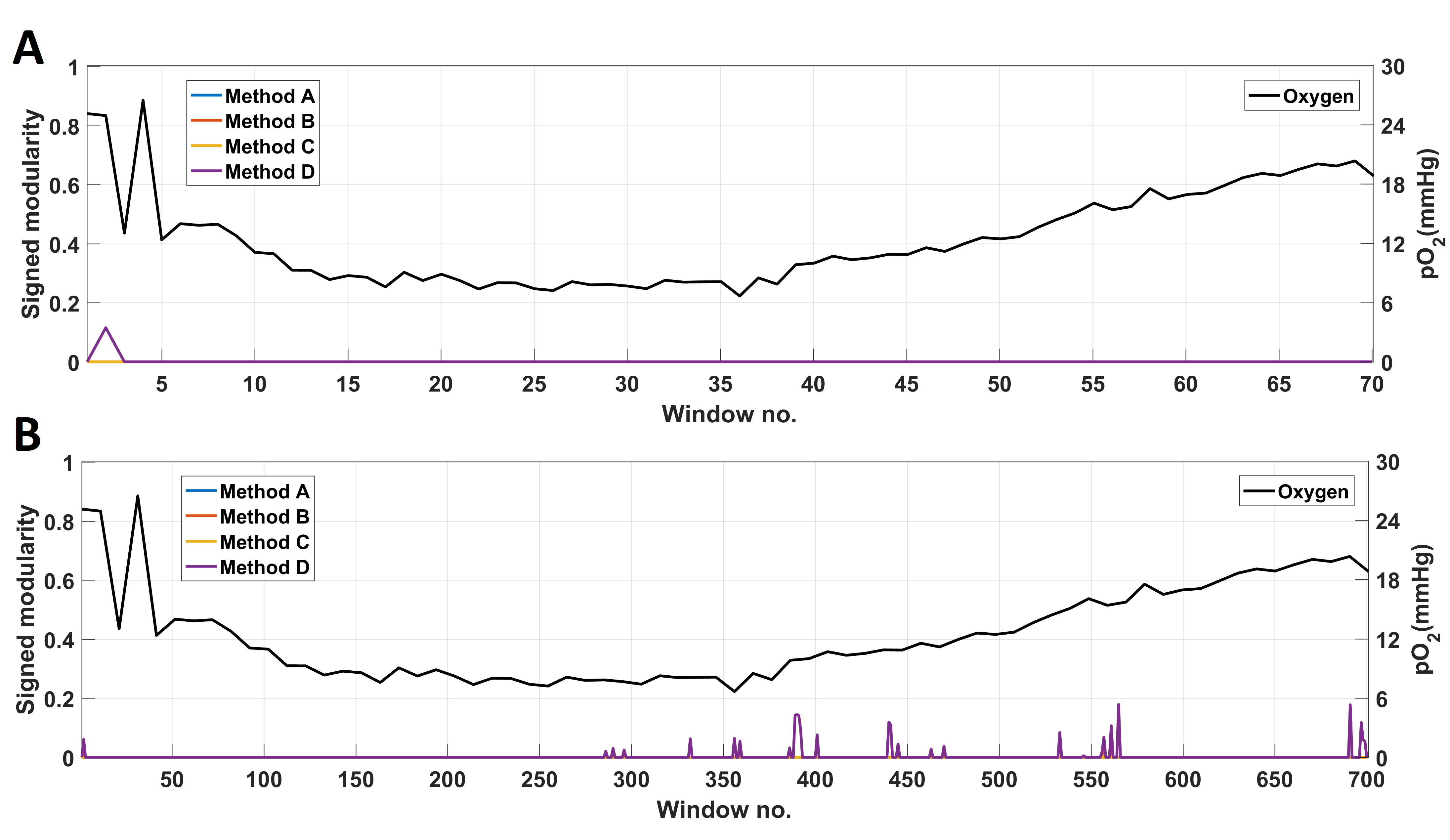}
		\end{center}
	\caption{The signed modularity obtained from all four methods for coherency matrices. A: window size: $100000$ and B: window size: $10000$.}
	\label{fig12}	
	\end{figure}

	 On the other hand, the results for the coherency matrices shown in \ref{fig10} reveal no clustering for almost all windows. The modularities obtained from these clustering are shown together in Fig \ref{fig11} and Fig \ref{fig12}, respectively for correlation and coherency matrices. All four methods show similar results with slight changes in the members of clusters and the corresponding modularities. The differences are expected since the methods are different in nature. The similarities for the correlation matrices however are striking, which can be interpreted as the clusters being segregated so clearly that all four methods capture them almost identically. The maximum signed modularity obtained for each windows size resides well above $0.7$ which shows a significant separation in the clusters. As mentioned before, the method C shows slightly lower signed modularity for almost all windows, which means the clusters found are not as segregated as the ones found by other methods. However, the number of calculations of method C is significantly less than those of other three methods, which means the accuracy is traded out for speed.
	 
	Furthermore, all four methods find only one cluster for almost all of the windowed coherency matrices, with the exception of a few windows that show two clusters. These two found clusters again consist of vertices $1,2,\ldots,9$ in one and $10,11,\ldots,16$ in the other, with vertex $10$ moving between the clusters at times. However, method C always finds one cluster to be the optimal number of clusters. Finally, the modularity of the clusters found for the coherency matrices always remains below $0.2$ which shows the insignificance of the clusters.
	
	Further, we computed the anticorrelation index for each windowed correlation matrix. Although all the anticorrelation indices remain well below $0.35$, the changes in them from a timed window to the next follow closely the changes in the modularities of those windows, as shown in Fig \ref{fig13}.

\section{Discussion}
\begin{figure}[h]
		\begin{center}
			\includegraphics[width=\linewidth]{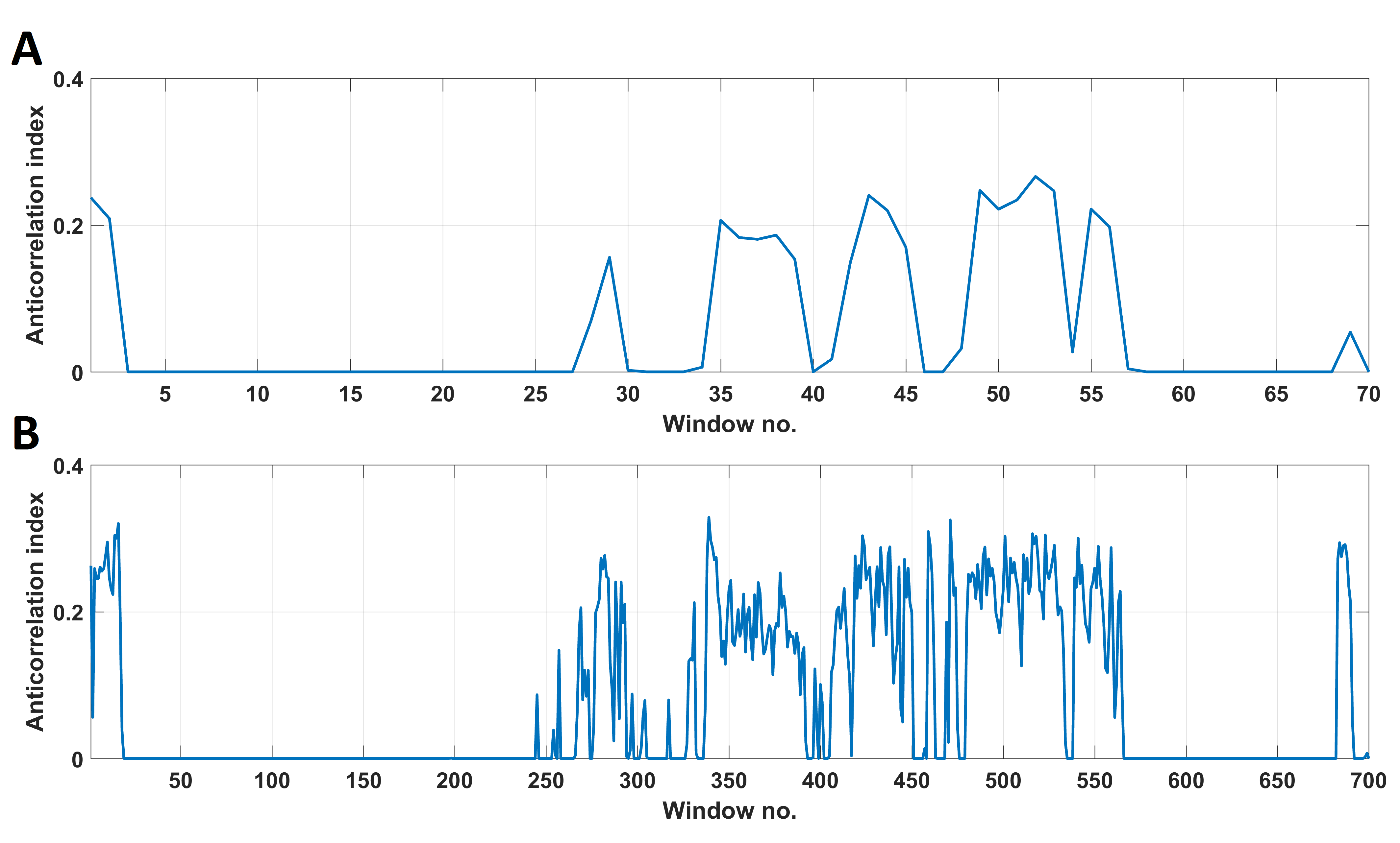}
		\end{center}
	\caption{The anticorrelation index for all windowed matrices of D$03$. A: window size: $100000$ and B: window size: $10000$.}
	\label{fig13}	
	\end{figure}

	Zero-lag correlation matrices or functional connectivity matrices have been the 
	source of community structure detection in both resting-state and task-oriented 
	BOLD fMRI data and in LFP data \cite{bullmoresporns09, gadelkarimetal14, 
	gadelkarimetal12, meunierlambiottebullmore10}. Here, we chose to study the community 
	structure among the LFP's recorded at a sampling rate of $1000$Hz in the CA1 
	region of the hippocampus of rats to study the pre-ictal and post-ictal 
	states. In particular, we investigated the LFP's for an anesthetized rat, D$03$, following an induced seizure which showed that there 
	are changes in the modular structure during the recorded time. Changes in some functional connectivity 
	matrices correspond to signed graphs, and reflect in some off-diagonal elements being 
	negative. This, as noted earlier can be calculated as the anticorrelation index of the matrix (ration of total of negative weights to the total of abosulte values of all weights) depicted in Fig \ref{fig13}. The changes in the signed modularity of the correlation matrix closely follows the changes of the anticorrelation index of it. This brings up the question about what role the anticorrelation of a signed graph plays in clustering it.
	
	Fig \ref{fig7} also reveals changes in the modular 
	structure of functional connectivity matrices. Since the CA1 region of the hippocampus 
	is a smaller region in the brain, we examine here the hierarchical organization of the 
	modular structure at a small scale. We found answers to the modularity structure of 
	the CA1 region under the condition described using four different unsupervised clustering 
	methods, the Fiedler 
	method \cite{fiedler75}, the spectral coordinates method \cite{wuwululi17}, the 
	Girvan-Newman method \cite{girvannewman02}, and the simulated annealing on signed modularity method 
	\cite{gomezetal09, gadelkarimetal14}. Fig \ref{fig11} 
	summarizes the calculated modularities and the corresponding number of clusters with the methods. 

	Four different methods used here have been discussed separately in literature 
	\cite{fiedler75, wuwululi17, newmangirvan04, gadelkarimetal14} . Because of the 
	significance of signed graphs either in the eigenvalue spectrum or in the non-linear 
	dynamics even in the single neuronal sense, a comparative study such as the one 
	presented here is necessary. Results in the case of an induced seizure on an 
	anesthetized rat (D$03$) show similar modularity 
	structure (Fig \ref{fig11}) in shape but certain differences among the methods. 
	What is important is that the four different
	methods show dramatic changes in the community structure at and immediately
	after the on-set of seizure with a concomitant change in
	going from the normoxic state to the hypoxic state. The hypoxia to normoxia
	period is a long duration one showing an upward ramp structure. The ramp in oxygen profile (See Fig \ref{fig3}) is punctuated with certain fluctuations that are in coincidence with
	community structure changes as reflected in the changes in the modularity
	index values. 
	Also, considering the number of clusters in an optimal clustering, we find that it shows fluctuations at different time windows
	similar to the fluctuations in computed modularity. However, there are slight
	differences in the number of modules or clusters the functional connectivity
	matrices among methods can be decomposed into one, two, or more clusters. We have
	not investigated the sensitivity of the maximal modularity index calculated to
	slight changes in the number of clusters. While the main focus in this paper is
	how certain mathematical methods are applied to this intriguing
	data related to normoxia-hypoxia-normoxia transitions in one anesthetized rat (D$03$),
	we need to be careful in drawing immediate
	conclusions.
	
	An examination of clustering, modularity index, and anticorrelation index results for two different windows sizes, $100000$ samples and $10000$ samples, as given in Figs \ref{fig9}, \ref{fig10}, \ref{fig11}, \ref{fig12}, and \ref{fig13}, shows that the smaller window size brings out highly-resolved features, suggesting a smoothing effect on them when one uses larger window sizes.

\section{Conclusions}
	Community structure in brain networks plays an important
	role in understanding brain function. We evaluated it in time-evolving
	networks from the electroencephalograph recordings in the CA1 hippocampal
	regions of an anesthetized rat. One crucial property of the community
	structure is modularity. We used four different methods, namely, the Fiedler
	method, the spectral coordinates method, the Girvan-Newman method, and
	the simulated annealing method to maximize the signed modularity. We
	conclude that the modularity index maps derived from the four methods used
	show similar community structure. We also find that the signed modularity
	index of community structures is more pronounced than the modularity index
	map of their unsigned counterparts, demonstrating the role of anticorrelation.

\section*{Acknowledgments}
	The authors would like to acknowledge the financial support of Canadian Institutes 
	of Health Research and Natural Sciences and Research Council of Canada. The authors 
	express deep gratitude to Dr. Michael Cavers, the department of Mathematics and 
	Statistics and the Hotchkiss Brain Institute at the University of Calgary for useful 
	discussions and support. KHM would like to thank Pacific Institute of Mathematical 
	Sciences (PIMS) for a post-doctoral fellowship. We thank Clayton T. Dickson for use of the recording equipment. Some of the computations have used 
	the Matlab code provided in \cite{MITSTRATEGIC} that we would like to thank them for 
	openly distributing the code.

\bibliography{ref}

\end{document}